\documentstyle[psfig,aps,prb, multicol]{revtex}

\newcommand{\ba}{\begin{eqnarray}}                                             
\newcommand{\ea}{\end{eqnarray}}     

\newcommand{\be}{\begin{equation}}                                             
\newcommand{\ee}{\end{equation}}

\begin{document}

\title{A close look at Jeckelmann's weak-coupling results }

\author{G. P. Zhang}

\address{Department of Physics, Indiana State University, Terre Haute,
IN 47809}

\draft

\date{\today}

\maketitle

\begin{multicols}{2}

In his comment,\cite{comment} Jeckelmann claims that our results
\cite{prb} are inaccurate and our criticism of his work\cite{eric} is
groundless, but unfortunately, in contrast to his assertion, our
conclusion is firmly grounded on a well-established fact, independent
of our numerical calculation. The fact is that at the weak-coupling
limit, $U/V_c$ must approach to 2.  Jeckelmann's results deviate from
2 in the weak-coupling limit, and therefore are troublesome.  In his
Comment, while avoiding this basic fact, Jeckelmann does not provide
new information about his calculation but instead he tries to create
more confusions by making the same mistakes as his previous
paper\cite{eric} did, which is unhelpful to the undergoing debate on
the phase diagram.

On the other hand, our results agree with both the weak- and
strong-coupling limit results, thus our conclusions in our paper
\cite{prb} remain unchanged.  In the following, we will investigate
and reply to his comments in detail.

First, Jeckelmann completely ignored the first density-matrix
renormalization group (DMRG) calculation \cite{prbold} on the phase
transition in the extended Hubbard model.  As shown in our previous
papers,\cite{prb,prbold} since $V_c$ is so close to $U/2$, a $V_c$
versus $U$ plot or a variation of it is very insensitive to the phase
transition boundary.  Jeckelmann is not able to show a critical figure
($U/V_c$ versus $U$) that we plotted in our paper.\cite{prb} This is
the key mistake that he made before\cite{eric} and now in this
comment\cite{comment} again.  His Fig. 1 has the same problem as his
old figure has, which hides the important difference between his
results and the weak-coupling limit results though his Figure 1
already shows the deviation of his results (open diamonds) from the
weak-coupling limit.  This unfortunately misleads him to conclude his
results are in quantitative agreement with other works, but it is
interesting to note that his paper has already been commented by the
same group of researchers that Jeckelmann claims his results agree
with, (for details see Ref.\cite{cam}).  If he plotted his data on a
$U/V_c$ versus $U$ plot, he would immediately see a deviation from the
weak-coupling limit results. To be clear, in Fig. 1 we show such a
deviation of his results, which we will come back to below.

Second, Jeckelmann\cite{comment} repeatedly tries to claim his results
agree quantitatively with the formal quantum Monte-Carlo
results,\cite{campbell} which is untrue.  The reason is that he has
eight (8) sets of data while three-four (3-4) sets of data are
available in the work by Sengupta.\cite{campbell} We pointed out his
problem in our paper.\cite{prb} We do not understand why the author could
fail to see the difference between them.  Instead of directly
addressing our concerns, he questions our calculation.

Third, all the numerical methods have their limitations. We have been
familiar with those references for a long time though Jeckelmann
failed to notice our previous paper.\cite{prb} The results by Cannon
{\it et al.}  were obtained in small clusters.\cite{can} Naturally,
the finite-size effect is unavoidable. Hirsch's MC results\cite{hir}
were obtained at one fixed temperature $\tau$. Cannon {\it et al.}
showed that the MC results depend on $\tau$, thus it is necessary to
perform $\tau\rightarrow 0$ extrapolation in order to get accurate
results. Nakamura\cite{jp} pointed out that the direct CDW-SDW level
crossing point has a large size-effect for all regions.  As already
mentioned in our paper, our calculation, not different from all the
other DMRG calculations, has several errors of different kinds:
truncation error, finite-size effect, $U$-dependent error, numerical
truncation error and many others.  That is why we are very careful
when we presented our results. Our results are obtained for system of
40 sites, truncation of 400 states, without any extrapolation and
within the infinite-system algorithm.  Our results are different from
those above results since our calculations are done in longer chains,
have smaller finite-size effects, do not have problems with
temperature, and have a better control over state
truncation. Therefore, the results tend to be more accurate. In
particular, our results are fully consistent with both strong- and
weak-coupling limit results. The reason why our present results are
different from our previous results\cite{prbold} is simply because at
present we used larger systems and higher truncation, then the results
certainly become more accurate. In fact, in our paper we have
systematically investigated this change (see Fig. 2 of our paper
\cite{prb}), while there is no such result in Jeckelmann's paper.

Fourth, Jeckelmann mentioned the infinite-system algorithm is known to
give incorrect results {\it occasionally}, but this does not
necessarily mean the results by the infinite-system algorithm are
wrong.  Without providing any evidence, he asserts the failure of the
standard infinite-system DMRG algorithm for the present problem.  He
goes on to provide one example in his reference 9 of his comment to
justify his argument. We looked at the reference and found that paper
is about a Bosonic Hubbard model not about the extended Hubbard model.
While it is true that the finite-system algorithm is more accurate,
what is surprising is that his results from an algorithm that he
claims `exact' are not consistent with the weak-coupling limit
results.

Fifth, the contradiction of his results with the weak-coupling limit
is truly troublesome.  Now he admits that his investigation focused on
the intermediate coupling regime, but nowhere in his original letter
mentions this, nor does this help him explain away the
contradiction. As shown in our previous paper, the trend in his
results is incorrect and deviates from $U/V_c=2$ substantially (see
Fig. 1). Jeckelmann now bypasses this comparison by saying that the
$U\rightarrow 0$ of $U/V_c$ cannot be determined using the data
obtained in his work. In order to justify his claim, he cited the
paper by Nakamura\cite{jp} that the weak-coupling limit is recovered
only for $U$ much smaller than $2t$.  We checked Nakamura's
paper.\cite{jp} To our surprise, in sharp contrast to Jeckelmann's
claim, Nakamura\cite{jp} showed that the intermediate-coupling is at
$U=4t$ and smaller than $4t$ the results asymptotically approach to
the weak coupling limit (see Fig. 12 in Nakamura's paper\cite{jp}).
This further demonstrates Jeckelmann's results are troubling.  It is
our hope that the author should directly and seriously address those
issues since the evidence is very clear.

Finally, for some reason, Jeckelmann apparently misunderstood our
statement in our paper.\cite{prb} When we said that we could not find
enough information, we mean that we could not find enough information
in {\it his paper} not in the cited references. By contrast, we are
pretty familiar with all those references for a long time.  In his
paper there is no information about system size, truncation, details
about the extrapolation and how to determine the phase
boundaries. That is why we were not able to compare our results with
his results, which is a motivation of our study.  Once again, his new
Fig. 1 avoids the critical comparison among different calculations and
does not provide a useful checkup of his data. Finally, we would like
to take this opportunity to point out that in scientific community
words like `ludicrous' should be avoided and it is our belief that a
little criticism on one's work is very healthy for the author
himself/herself as well as for the entire community.

In conclusion, we emphasize that none of our criticism on his paper as
well as our main conclusion has been changed. We kindly suggest that
the author check his data and address the criticisms more directly and
throughly.

\begin{figure}
\caption{The critical ratio $U/V_c$ versus $U$ obtained in Ref. 4
demonstrates a strong deviation from the weak-coupling limit.  The
weak-coupling limit is highlighted by a box at the top.  }
\end{figure}

\vspace{1cm}\psfig{figure=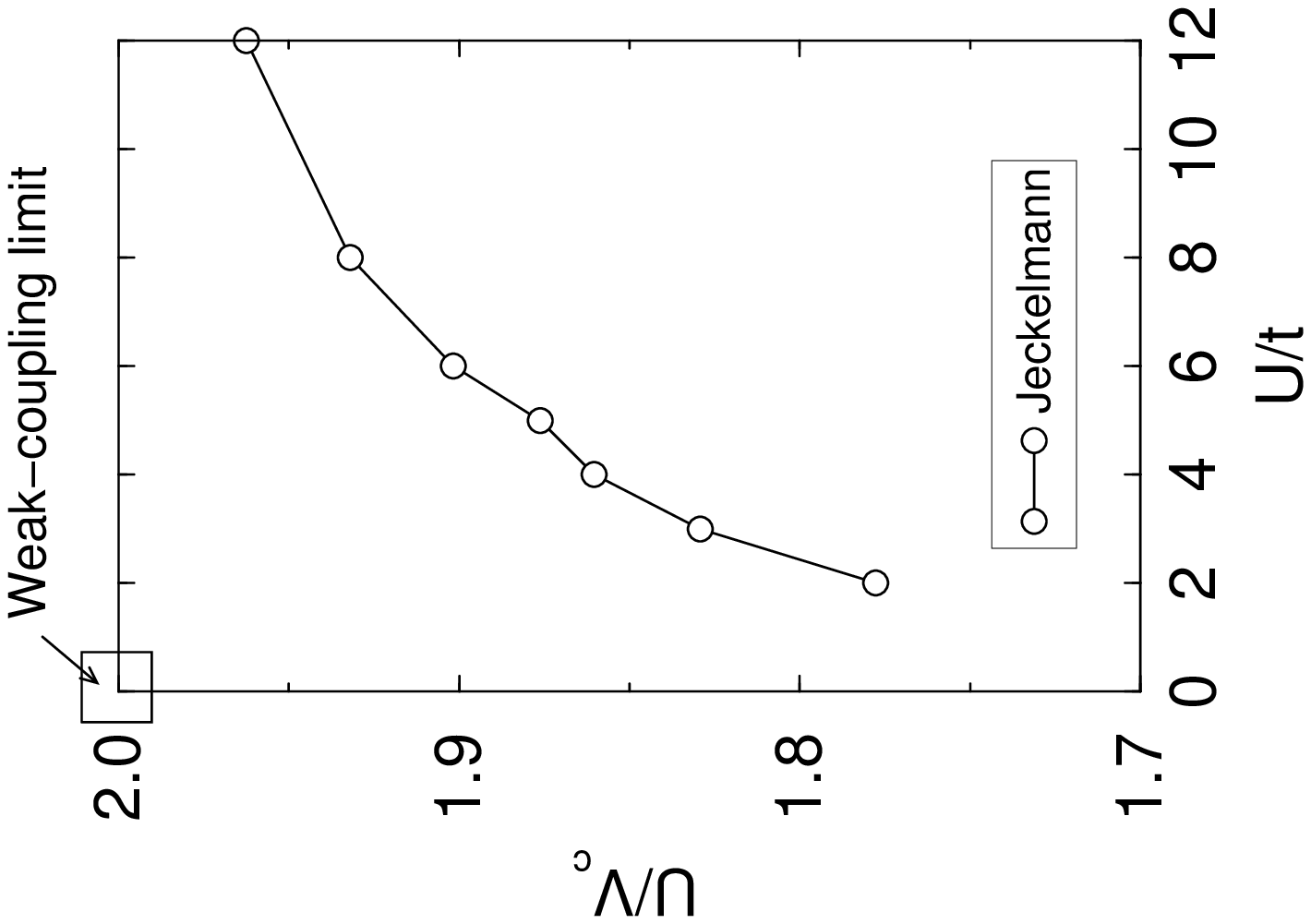,width=7cm,angle=270}

\end{multicols}

\end{document}